\pdfoutput=1

\documentclass[pra,twocolumn,superscriptaddress,nofootinbib]{revtex4}

\usepackage{amsmath,amsfonts,amssymb,amstext}
\usepackage[dvipdfm]{graphicx}
\usepackage[colorlinks,linkcolor=blue]{hyperref}
\newcommand{\bra}[1]{\ensuremath{\langle #1 \vert}}
\newcommand{\ket}[1]{\ensuremath{\vert #1 \rangle}}
\newcommand{\braket}[2]{\ensuremath{\langle #1 \vert #2 \rangle}}
\newcommand{\mbf}[1]{\mathbf{#1}}

\begin{document}
\title{Quantum ergodicity and entanglement in kicked coupled-tops}
\date{\today}

\author{Collin M. Trail}
\email{ctrail@unm.edu}
\affiliation{Department of Physics and Astronomy, University of New Mexico}
\author{Vaibhav Madhok}
\affiliation{Department of Physics and Astronomy, University of New Mexico}
\author{Ivan H. Deutsch}
\affiliation{Department of Physics and Astronomy, University of New Mexico}

\begin{abstract}
We study the dynamical generation of entanglement as a signature of chaos in a system of periodically kicked coupled-tops, where chaos and entanglement arise from the same physical mechanism.   The long-time averaged entanglement as a function of the position of an initially localized wave packet very closely correlates with the classical phase space surface of section -- it is nearly uniform in the chaotic sea, and reproduces the detailed structure of the regular islands.  The uniform value in the chaotic sea is explained by the random state conjecture. As classically chaotic dynamics take localized distributions in phase space to random distributions, quantized versions take localized coherent states to pseudo-random states in Hilbert space. Such random states are highly entangled, with an average value near that of the maximally entangled state. For a map with global chaos, we derive that value based on new analytic results for the typical entanglement in a subspace defined by the symmetries of the system.  
 For a mixed phase space, we use the Percival conjecture to identify a ``chaotic subspace" of the Hilbert space.  The typical entanglement, averaged over the unitarily invariant Haar measure in this subspace, agrees with the long-time averaged entanglement for initial states in the chaotic sea.   In all cases the dynamically generated entanglement is predicted by a unitary ensemble of random states, even though the system is time-reversal invariant, and the Floquet operator is a member of the circular orthogonal ensemble.
\end{abstract}

\maketitle

\section{Introduction}
The connections between complexity, nonlinear dynamics, ergodicity, and entropy production, have long been at the heart of the foundations of statistical physics. A central goal of ``quantum chaos" has been to extend this foundation to the quantum world.  Classic works on the subject including level statistics \cite{Berry77}, properties of Wigner functions \cite{Berry77a}, and quantum scars in ergodic phase spaces \cite{HellerScar} have tended to focus on the properties of wave mechanics, e.g. the dynamics of single particle billards \cite{McDonald} (also seen in the properties of classical waves, e.g. microwave cavities \cite{Stockmann}).  More recently, the tensor product structure of quantum mechanics, essential for understanding systems with multiple degrees of freedom, has come to the fore.  In that context, one is naturally led to consider how the dynamical generation of entanglement between quantum subsystems is connected with the chaotic dynamics of coupled classical degrees of freedom.  Such studies address fundamental issues of complexity in quantum systems and are potentially applicable in quantum information processing, where entanglement is considered to be an essential resource.  The goal of this article is to revisit this problem and establish a unifying theory that allows us to make quantitative predictions and gives qualitative understanding of the connection between chaos and entanglement.

The connection between chaos in the classical description of Hamiltonian dynamics and entanglement in the quantum description has been the subject of extensive study over the last decade.  The original motivation of Zurek and Paz  was to address the quantum-to-classical transition \cite{Zurek/Paz}.  By conjecturing that chaotic systems decohere exponentially fast through their entanglement with the environment, they hoped to resolve a paradox in which a macroscopic system would exhibit the effects of quantum coherence on a time scale logarithmic in $\hbar$.  Work quickly following this turned to studies of the coupling of just two degrees of freedom, rather than system-environment coupling, as entanglement is most easily quantified for bipartite systems  \cite{Fuyura}.  Considering two coupled kicked tops (a standard paradigm of quantum chaos \cite{Haake}) Miller and Sarkar \cite{Miller/Sarkar} correlated the rate of generation of entanglement with the Lyapunov exponents associated with the mean positions of quantum wavepackets localized in phase space.

In related research, Bandyopadhyay and Lakshminarayan \cite{Lakshminarayan/Bandyopadhyay2002,Lakshminarayan/Bandyopadhyay2004} explored the amount of entanglement that is associated with coupled kicked tops, with particular emphasis on the entanglement of the Floquet eigenstates \cite{Lakshminarayan}.  The entanglement of these eigenstates saturated to a value below the maximum possible value in a way that depended only on the Hilbert space dimension, not the chaoticity parameter.  The same was was true of the dynamically generated entanglement.   This work gave the first indication that the entanglement generated by the coupled tops was statistical in nature, and related to the theory of random states in Hilbert space. Using random matrix theory \cite{Mehta, Haake} they were able to determine the statistics of the Schmidt coefficients of a random bipartite pure state, and thus were able to predict the saturation value of the entanglement for the Floquet eigenstates. 

The relationship between the entanglement in the eigenstates and the dynamically generated entanglement is subtle \cite{Kus04}; we'll return to this point later.  Ghose and Sanders have shown that there are signatures of chaos in the dynamically generated entanglement in a single kicked top when the large angular momentum is thought of as a collection of symmetrically coupled qubits \cite{Ghose, Wang}.  They used the Floquet spectrum to explain the initial rise time and power spectrum in the entanglement history.  Dynamical generation of entanglement by chaotic maps, and its relation to random unitary matrices was also explored  by Scott and Caves \cite{Scott/Caves} as a way of comparing different quantizations of the Baker's map, and by Viola and coworkers as a means of quantifying complexity in quantum systems and its relationship to generalized entanglement \cite{Viola}.

While many of the elements connecting chaos and entanglement have been explored with a variety of successful numerical and approximate analytic predictions, there has been no unifying theory, and in some cases, the key relations have been obscured.  The most extensive studies, discussed above, consider coupled systems which are separately chaotic (kicked tops) rather than being chaotic due their coupling.  This can muddy the waters.  For example, it was shown that the strength of chaos did not enhance the production rate of entanglement for weakly chaotic coupled tops \cite{Fujisaki}, while Bandyopadhyay and Lakshminarayan \cite{Lakshminarayan/Bandyopadhyay2002,Lakshminarayan/Bandyopadhyay2004} showed that the relationship between entanglement and the chaoticity parameter varied with the coupling strength between the tops.

 As we seek to connect chaos and entanglement, it is most natural to consider systems in which they arise from the {\em same mechanism} -- their physical coupling.  Moreover, by considering a system that is chaotic only when the two parts are coupled, the classical phase space describes the global system rather than a subsystem, and there is no ambiguity about the nature of the dynamics.  The distinction between weak and strong coupling cannot be made independent of strong and weak chaos, thereby sharpening our focus on the key relationships. 

To address these issues, we consider a model system of kicked coupled-tops, rather then coupled kicked-tops, described in detail in Sec. II.  This system is motivated by its connection to possible experimental realizations, our ability to easily visualize the classical phases space, and to analyze the Floquet map.  We use this system as the forum to explore the basic thesis of this paper.    Chaos arises in classical dynamics because of insufficient symmetry (integrals of motion) for a given number of degrees of freedom.  In the quantum analog, insufficient symmetry leads to the random matrix conjecture -- systems with global classical chaos have eigenvectors and eigenvalues that are statistically predicted by ensembles of random matrices \cite{Bohigas, Haake}.  Moreover, whereas global chaos leads to ergodic dynamics and the generation of ``random" coarse-grained distributions on phase space, a large body of numerical studies indicate that the quantum chaotic map is ergodic in the sense that it generates a state with many properties that are statistically predicted by a random state in Hilbert space, picked according to appropriate Haar measure \cite{Scott/Caves}.  The dynamically generated entanglement is then that of a random state (by this measure) in the relevant Hilbert state.  These predictions can be extended to mixed phase spaces with regular islands immersed in a chaotic sea.  With the help of Percival's conjecture \cite{Percival} that divides eigenstates into chaotic and regular classes, ergodicity on the chaotic sea leads to random states in a chaotic subspace and a commensurate typical entanglement.  Whereas in the globally chaotic case we can derive analytic results, for the mixed phase spaces we are relegated to numerical predictions, which nonetheless verify the ergodic conjecture, connecting entanglement generation in chaotic dynamics to the creation of pseudo-random states in Hilbert space.

The remainder of this paper is organized as follows.  In Sec. II we introduce our model of kicked coupled-tops, studying the classical and quantum features. Section III, the heart of paper, studies the entanglement in our system.  We perform numerical calculations of the entanglement of the system's eigenstates, the long-time averaged entanglement generated by the Floquet map, and its relationship to the classical phase space.  We then explain these results in terms of the properties of random states in Hilbert space.  Reviewing the essential ideas, we derive new analytic expressions for the typical entanglement of a random state when we are restricted to a subspace of the full tensor product space.  This is of relevance here given the symmetries of the system.  We also pay particular attention to the subtle distinctions between the eigenstates of random matrices and the random states generated from initially localized wavepackets.  In doing so we clarify previous works and make accurate predictions, especially for global chaos, but also extended to a more general mixed phase space scenario.  Our results are discussed and summarized in Sec. IV.

\section{Kicked Coupled-Tops}
\subsection{Quantum and classical descriptions }
We consider a bipartite system composed of two spins, $\mbf{I}$ and $\mbf{J}$,  isotropically coupled in a Heisenberg interaction, and subject to periodic kicks that act only on spin $\mbf{J}$. Choosing the direction of the kicks to be about the $z$-axis, the system evolves according to the Hamiltonian,
 \begin{equation}
 H=A \mbf{I} \cdot \mbf{J}+\sum_{n=-\infty}^\infty \delta (t - n \tau) B J_z .
 \end{equation}
Here $A$ gives the strength of the isotropic coupling, $B$ the strength of the kicking, and $\tau$ is the kicking period. Such a Hamiltonian describes the hyperfine interaction between nuclear spin $\mbf{I}$ and total electron angular momentum $\mbf{J}$, with a magnetic field that has negligible effect on the nucleus.  While this realization cannot reach deep into the semiclassical regime, for large atoms, with heavy nuclei and a large number of electrons in the valance shell, one can explore nontrival mesoscopic regimes.  The true semiclassical limit can potentially be attained in an atom-photon system where $\mbf{I}$ is the collective spin of an atomic ensemble coupled to the Stokes vector $\mbf{J}$ of a quantized electromagnetic field \cite{Mitchell}.  We will not consider here the feasibility of experimental realizations, instead focusing on the foundational theory.

\begin{figure}
\includegraphics[width=\linewidth]{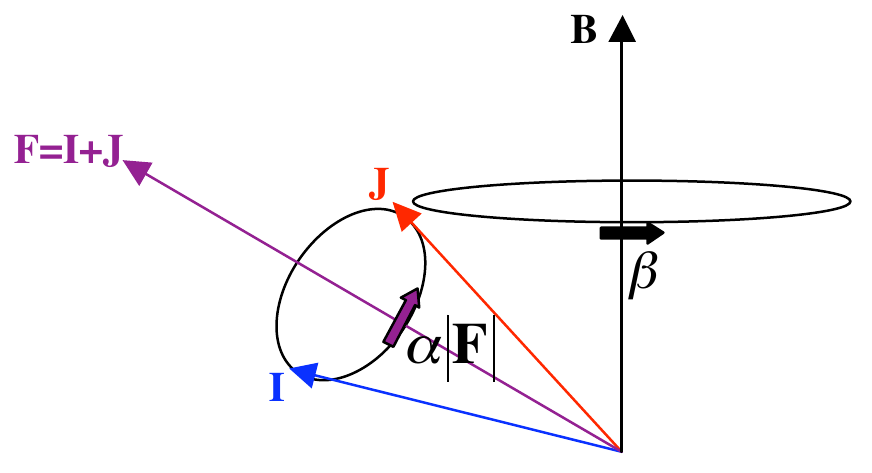}
\caption{The dynamics of the kicked coupled tops can be viewed as an alternating sequence of rotations. The two spins $\mbf{I}$ and $\mbf{J}$ precess around the total angular momentum $\mbf{F}$ by an angle $\alpha |\mbf{F}|$, and the spin $\mbf{J}$ is kicked around the space-fixed $z$-axis by $\beta$.}
\label{F1}
\end{figure}

Choosing the external field to act in delta kicks allows us to express the Floquet map (transformation after one period) in a simple form of sequential rotations,
\begin{equation}
\label{Eq:Floquet}
U_{\tau} = e^{-i\alpha \mbf{I} \cdot \mbf{J}}  e^{-i \beta J_z} \equiv e^{-i\alpha F^2 /2}  e^{-i \beta J_z},
\end{equation}
where $\alpha$ and $\beta$ are related to $A$ and $B$ in terms of the kicking period, $\hbar$, etc.  In the second form, we have expressed the rotation in terms of the total angular momentum $\mbf{F} = \mbf{I} + \mbf{J}$ and neglected irrelevant overall phases.  We can thus interpret the dynamics as alternating a rotation of $\mbf{J}$ about a space fixed $z$-axis by angle $\beta$, followed by a procession of $\mbf{I}$ and $\mbf{J}$ about $\mbf{F}$ by an angle $\alpha |\mbf{F}|$, as shown in Fig (\ref{F1}).  Such a simple transformation nonetheless leads to complex dynamics, including chaos in the classical limit as discussed below.  From the quantum perspective, since the two rotations don't commute, there are insufficient symmetries to specify Floquet eigenstates by a complete set of commuting operators; the system is not integrable.  Note, however, that the system is invariant under an overall rotation around the $z$-axis, so $F_z$ is a conserved quantity ($F^2$ is not conserved).

We treat the classical limit of quantum mechanical spin in the familiar way \cite{Haake}. Each of our spins has three components, but a fixed magnitude, and thus their orientations can be specified by two variables. The $z$-component of a spin and the angle $\phi$, denoting its orientation in the $x-y$ plane, are canonically conjugate, and thus each spin constitutes one canonical degree of freedom.  The classical dynamical map has the same physical action as described above in the quantum context -- rotation of $\mbf{J}$ by angle $\beta$ followed by precession of $\mbf{I}$ and $\mbf{J}$ about $\mbf{F}$ by angle $\alpha |\mbf{F}|$.  Here, the rotations are implemented by $3 \times 3$ SO(3) matrices.  The two spins, plus time-dependent Hamiltonian imply a five dimensional phase space.  Since $F_z$  is conserved, the dynamics is restricted to a four-dimensional hypersurface.  As there are no additional constraints, the dynamics are not integrable and can exhibit chaos.

To visualize the dynamics, we rewrite our system in terms of a new set of variables, $(F_z, \bar{\phi} \equiv \phi_I+\phi_J)$ and $(\delta F_z  \equiv I_z-J_z, \delta\phi \equiv\phi_I-\phi_J)$, 
 \begin{subequations}
  \begin{align}
J_z &= \frac{F_z-\delta F_z}{2}, \\
\mbf{I} \cdot \mbf{J} &= I_z J_z + IJ \left( \sin \phi_I \sin \phi_J+\cos \phi_I \cos \phi_J \right) \nonumber \\
&= \left( \frac{F_z+\delta F_z}{2}\right) \left( \frac{F_z-\delta F_z}{2} \right) +IJ\cos(\delta \phi).
 \end{align}
\end{subequations}
Because $F_z$ is a conserved quantity, $\bar{\phi}$ does not appear in our Hamiltonian. It is a cyclic coordinate, and thus we can ignore it without losing any information about the further evolution of the remaining variables. Neither do we require $\bar{\phi}$ to determine the Lyapunov exponent of a chaotic system. Thus, we need only consider the two difference variables,  $(\delta F_z , \delta\phi)$,  and time, taking us from a four to a three dimensional hypersurface. This allows us to visualize our system using a Poincar\'{e} surface of section as a stroboscopic plot.  We restrict our attention here to $F_z=0$ as this also leads to the largest subspace in the associated quantum problem.

\begin{figure}
\includegraphics[width=\linewidth]{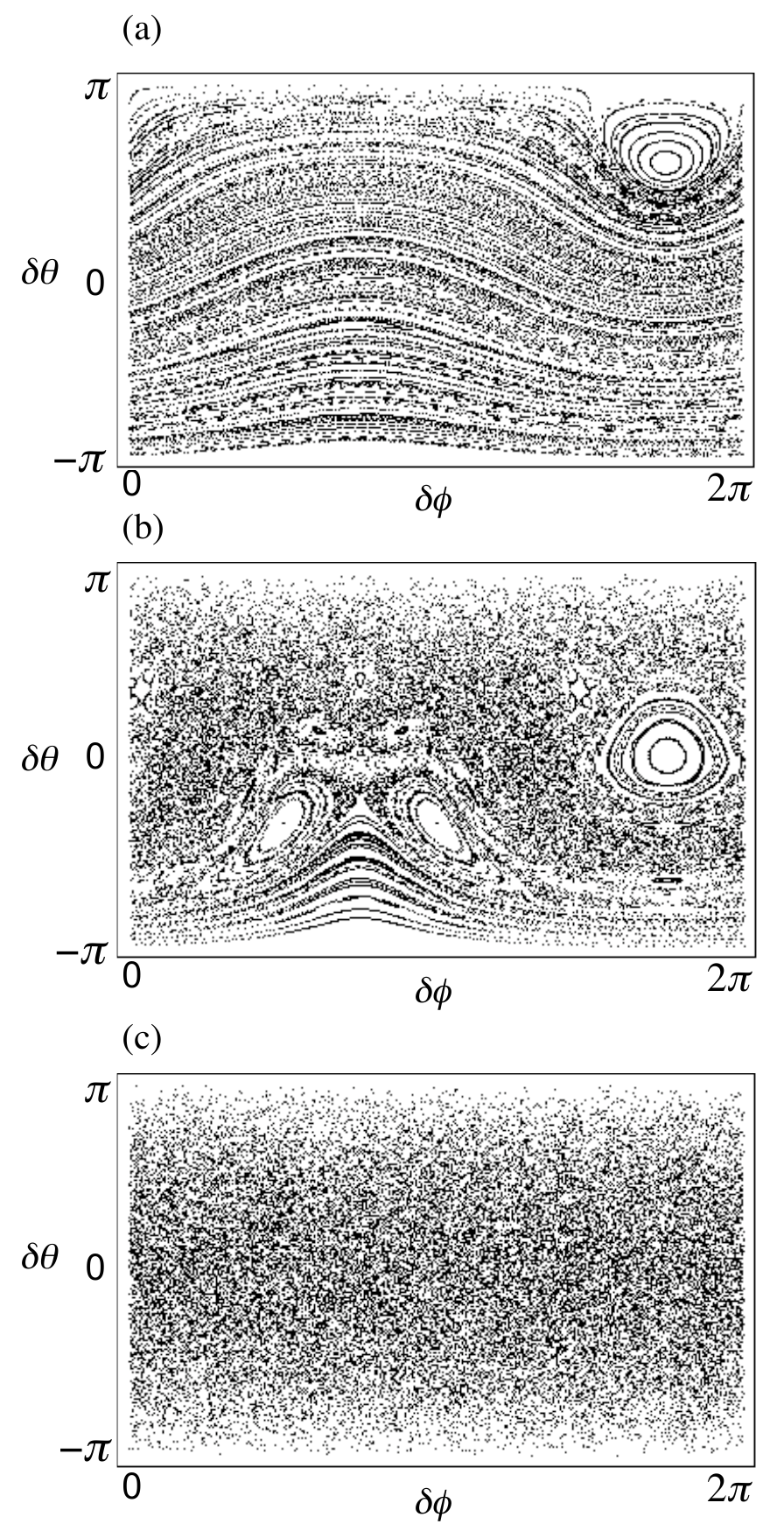}
\caption{Poincar\'{e} surface of section for the coupled kicked tops, with $F_z=0$ (a) Regular phase motion: $\alpha=1/2,\beta=\pi/2$, (b) Mixed phase space: $\alpha=3/2,\beta=\pi/2$, (c) Global chaos: $\alpha=6,\beta=\pi/2$.}
\label{F2}
\end{figure}

The classical equations of motion depend on the ratio $|\mbf{I}|/|\mbf{J}|$.  We focus here on equal spin magnitudes and fix $F_z=0$.  Thus, without loss of generality, since the SO(3) rotation matrices of classical dynamics are independent of spin magnitude, we take the spin vectors to be unit vectors. The basic structure of the phase space can be understood as follows. When the coupling is removed, our system has fixed points at the northern and southern ``poles". As the chaoticity parameter is turned up, chaos first forms around the unstable ``north pole" while regular behavior persists around the stable ``south pole". Further fixed points appear in the usual manner as bifurcations occur with increase of the chaoticity parameter. Figure (\ref{F2}) shows three different regimes of classical dynamics.  With the parameters $\alpha=1/2, \beta=\pi/2$ (Fig.  (\ref{F2}a), the dynamics are highly regular, with negligible stochastic motion.  When $\alpha=3/2, \beta=\pi/2$ (Fig.  (\ref{F2}b), we see a mixed space with chaotic and regular regions of comparable size. The parameters $\alpha=6, \beta=\pi/2$ (Fig.  (\ref{F2}c), give a completely chaotic phase space. 

We want to choose our quantum Hamiltonian so that we will recover our classical dynamics in the large spin limit. We would like to be able to vary the size of our spins, but we will keep the pair equal to each other in magnitude, $I=J$. Since the SU(2) rotation matrices depend on the spin magnitude, we must scale the Floquet operator.  By substituting $\alpha \rightarrow \tilde{\alpha}=\alpha/J$ we obtain the same Heisenberg equations of motion as the classical equations for equal magnitude spins.

\subsection{Quantum chaology}\label{S3}
In order to understand the dynamical generation of entanglement, we need to establish some basic understanding of the eigenstates of the system and their relationship to the classical dynamics.  As our system is time periodic, the states of interest are the eigenstates of the Floquet operator, Eq. (\ref{Eq:Floquet}).  It is useful to consider both the coupled and uncoupled representations of angular momentum connected by the usual Clebsch-Gordan expansion,
\begin{equation}
\ket{F,M_F} = \sum_{m_I,m_J}  \braket{F,M_F}{I, m_I; J,m_J} \ket{I, m_I}\ket{J,m_J}
\end{equation}
Conservation of $F_z$ implies that the operator is block diagonal for all states defined by quantum number $M_F$. The largest block, $M_F=0$, has dimension $2J+1$ as $F$ varies from 0 to $2J$. Using the uncoupled representation, denoting the product state by the single quantum number $m_J=-m_I$, the matrix,
\begin{eqnarray}
\bra{m'_J} U_{\tau} \ket{m_J}& =& \sum_{F} e^{-i \left( \alpha \frac{F(F+1)}{2J}+ \beta m_J \right) } \\
&  &\braket{F,0}{I, -m'_J; J,m'_J}  \braket{F,0}{I, -m_J; J,m_J} \nonumber 
\end{eqnarray}
can then be diagonalized to yield the Floquet eigenstates and eigenphases,
\begin{equation}
\label{Eq:Eigen}
\ket{k} = \sum_{m_J} c^{(k)}_{m_J}  \ket{I, -m_J}\ket{J,m_J}; \hspace{1 pc}
U_{\tau}\ket{k} = e^{-i \phi_k} \ket{k}.
\end{equation}

A central result of quantum chaos is the connection with the theory of random matrices \cite{Haake}.  In the limit of large Hilbert space dimensions (small $\hbar$), for parameters such that the classical description of the dynamics shows global chaos, the eigenstates and eigenvalues of the quantum dynamics have the statistical properties of an ensemble of random matrices.  The appropriate ensemble depends on the properties of the quantum system under time-reversal \cite{Haake}.  We thus seek to determine whether there exists an anti-unitary (time reversal) operator $T$ that has the following action on the Floquet operator,
\begin{equation}
T U_{\tau} T^{-1} = U_{\tau}^{\dagger} = e^{i \beta J_z}e^{i \tilde{\alpha} \mbf{I}\cdot \mbf{J}}.
\end{equation}
Analogous to the case of the single kicked top, we consider the generalized time reversal operation,
\begin{equation}
\label{Eq:Reversal}
T=e^{i \beta J_z} K,
\end{equation}
where $K$ is complex conjugation in the uncoupled product representation.  Since {\em both} $I_y$ and $J_y$ change sign under conjugation, while the $x$ and $z$ components do not,
\begin{equation}
K J_z K = J_z; \hspace{1 pc} K\mbf{I} \cdot \mbf{J}K = \mbf{I} \cdot \mbf{J}.
\end{equation}
It then follows that 
\begin{eqnarray}
T U_{\tau} T^{-1} &=& \left( e^{i \beta J_z} K \right) \left( e^{-i\tilde{\alpha} \mbf{I} \cdot \mbf{J}}  e^{-i \beta J_z}  \right) \left( K e^{-i \beta J_z} \right)  \\
&=&  e^{i \beta J_z} \left( e^{i \tilde{\alpha} \mbf{I}\cdot \mbf{J}} e^{i \beta J_z}\right) e^{-i \beta J_z} \nonumber\\
&=& e^{i \beta J_z}e^{i \tilde{\alpha}\mbf{I}\cdot \mbf{J}}= U_{\tau}^{\dagger}, \nonumber
\end{eqnarray}
so the dynamics are time-reversal invariant.  Moreover, $T^2=1$, so there is no Kramer's degeneracy. Given these facts, for parameters in which the classical dynamics are globally chaotic, we expect the Floquet operator to have the statistical properties of a random matrix chosen from the circular orthogonal ensemble (COE).

To further correlate the Floquet eigenstates with the classical phase space in the case of regular and mixed dynamics, it is useful to employ a Husimi representation. A spin coherent state has a minimum quantum uncertainty and is specified by polar orientation angles $\theta$ and $\phi$ on the sphere. In terms of the standard basis, a spin coherent state for a single spin is \cite{Puri}
\begin{equation}
\ket{\mu} = \sum_m \frac{\mu^{J-m}}{\left(1+|\mu|^2 \right)^J} \sqrt{\frac{(2J)!}{(J-m)!(J+m)!}}\space \ket{J,m},
\end{equation}
where $\mu =\tan(\theta/2) e^{i \phi}$.  For our system, because the subspaces in which the eigenstates live are not described by an irreducible representation of angular momentum, there are no such minimum uncertainty states for the difference angles.  Nonetheless, we obtain a useful set of states by projecting the product of spin coherent states associated with the two subsystems onto the subspace with a fixed value of $F_z$ (here $F_z=0$).  The result of the projection is
\begin{equation}
\label{Eq:Project}
\hat{P}_0 \ket{\mu_I} \ket{\mu_J} = \sum_m \left( \frac{\mu_I}{\mu_J} \right)^m \frac{(2J)!}{(J-m)!(J+m)!} \ket{m}_I \ket{-m}_J.
\end{equation}
Classically, in projecting on to the surface of section with $F_z=0$, we take $\theta_I + \theta_J = \pi$. Fixing this value in the quantum state one finds
\begin{equation}
\frac{\mu_I}{\mu_J}=e^{i(\phi_I-\phi_S)} \left[ \frac{1 + \sin\left( \frac{\theta_I - \theta_S }{2} \right)}{1 - \sin\left( \frac{\theta_I - \theta_S}{2}\right) }\right].
\end{equation}
The projected coherent state thus depends only on the difference of the angle variables, and allows us to consider localized quantum states correlated with the classical phase space of interest.  After normalizing, we arrive at an over-complete basis of states for the $F_Z=0$ subspace, parameterized by $\delta\theta$ and $\delta\phi$ . The Husimi distribution of a state $\ket{\psi}$ in this space,

\begin{equation}
Q(\delta\theta,\delta\phi) \equiv |\braket{\delta\theta,\delta\phi}{\psi}|^2
\end{equation}
then provides a visualization in phase space.
\begin{figure}
\includegraphics[width=\linewidth]{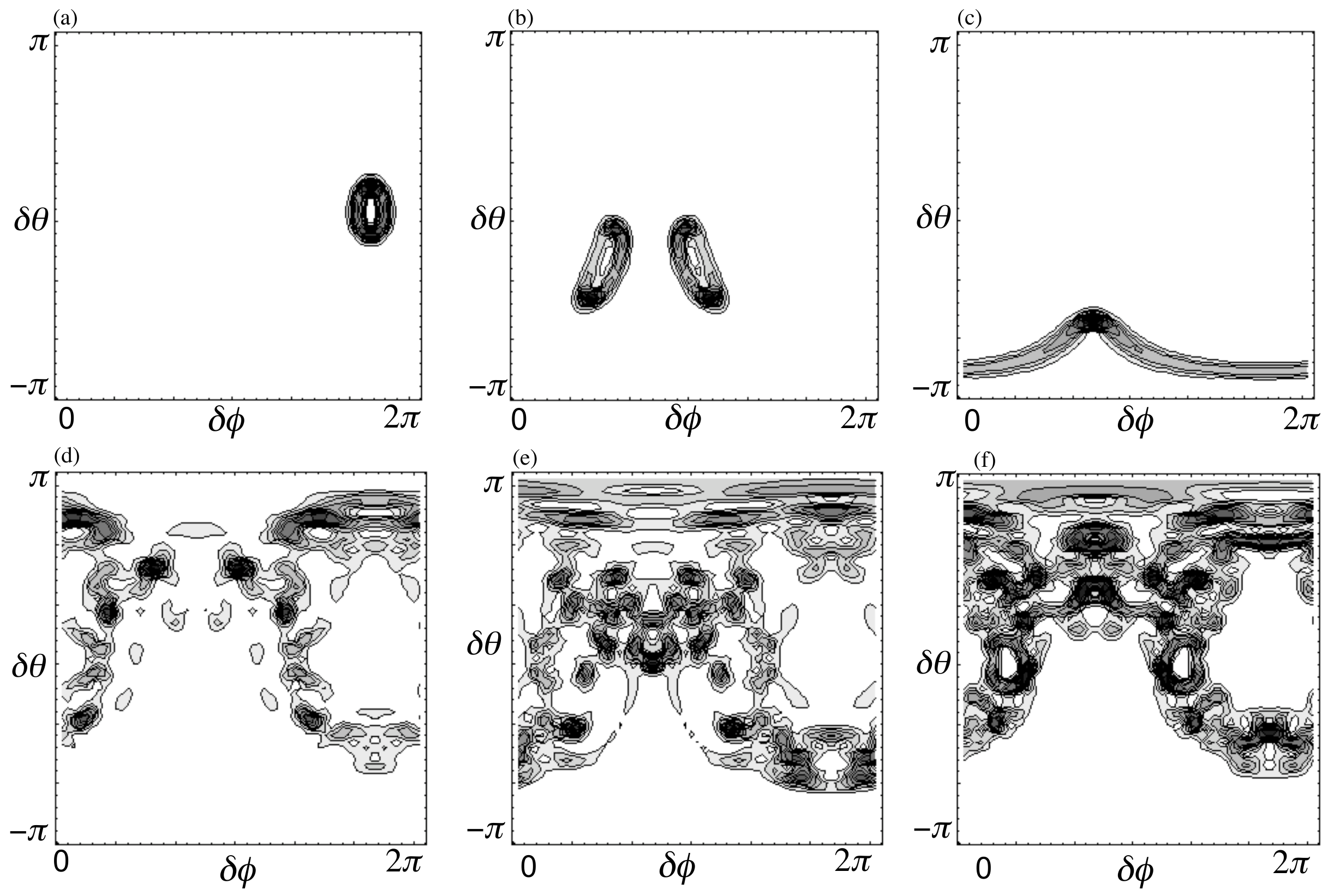}
\caption{ Husimi distributions of Floquet eigenstates associated with the parameters of a mixed phase space (Fig. \ref{F2}b). (a,b,c) Regular eigenstates around different fixed points. (d,e,f) Chaotic eigenstates, delocalized in the chaotic sea.}
\label{F3}
\end{figure}

In order to explore the semiclassical limit, we choose $I=J=150$, corresponding to a $d=301$ dimensional Hilbert space in the $F_z=0$ subspace, or an ``effective $\hbar$" of $\hbar_{eff} = 1/301$.  Figure \ref{F3} shows the Husimi plots of a few of the eigenstates for $\alpha/J=3/2, \beta=\pi/2$, for which the classical phase space is mixed (Fig. \ref{F2}b).  These plots exhibit the features expected according to Percival's conjecture.  The states roughly divide into regular and irregular sets, with regular eigenstates concentrated on invariant tori around stable fixed points, resembling harmonic oscillator eigenstates, and irregular ``chaotic" states randomly distributed within the chaotic sea.

\begin{figure}
\includegraphics[width=\linewidth]{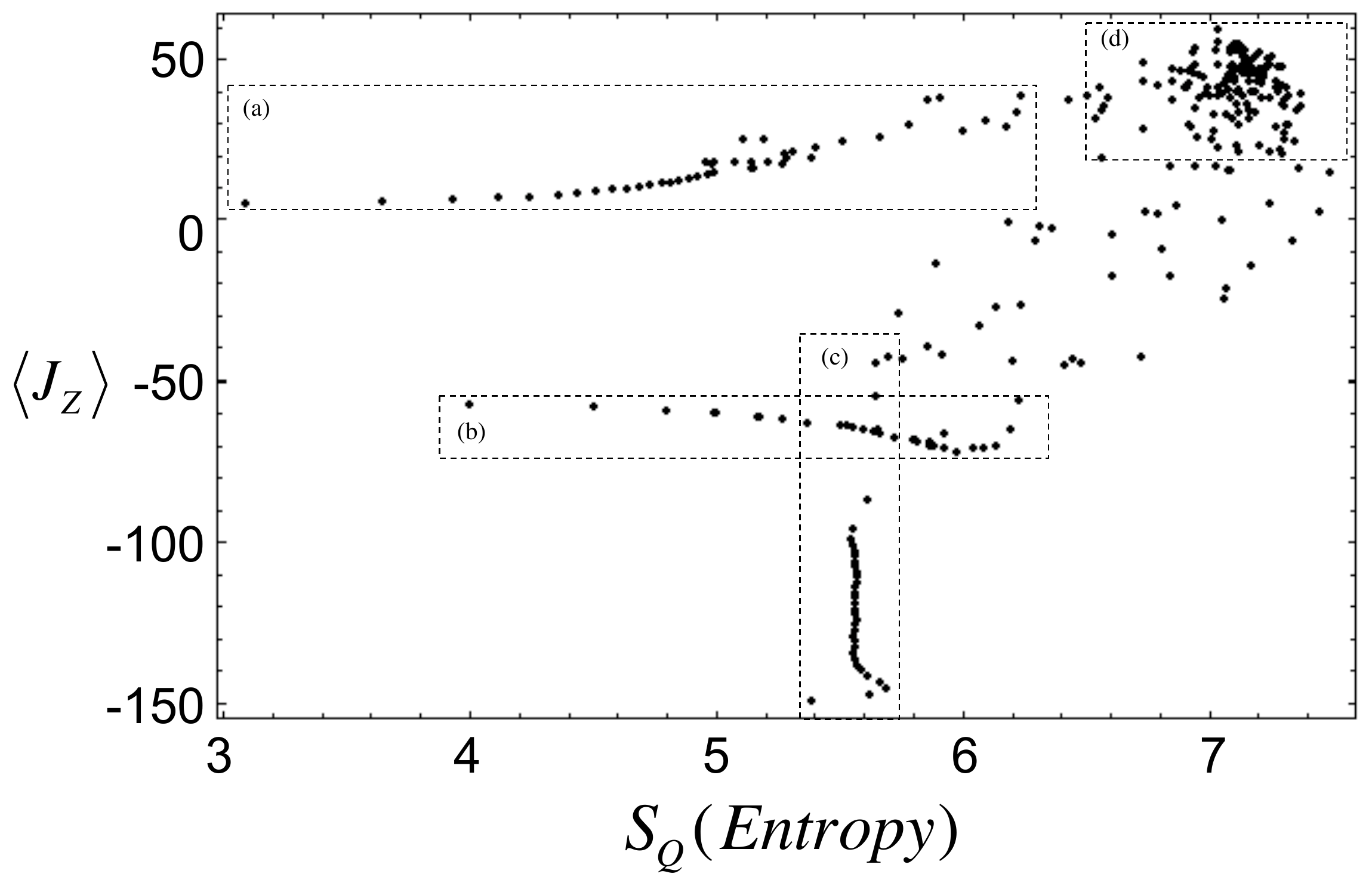}
\caption{Scatter plot of $\langle J_z \rangle$ vs. Husmi Entropy,  $S_Q$, for the Floquet eigenstates associated with the mixed phase space (Fig. (\ref{F2}b)). Boxed regions (a), (b), and (c) correspond to regular states centered around fixed points. States in region (d) are considered ``chaotic eigenstates".}
\label{F4}
\end{figure}

Though Percival's conjecture is largely born out in numerical analyses, it is not strictly true (especially in the finite $\hbar$ limit), nor is there a strict procedure for filtering the regular from chaotic eigenstates except for very special systems \cite{Tomsovic}. We can, nonetheless, create an approximate filter.  A useful measure for distinguishing states is the Shannon entropy of the Husimi distribution, 
\begin{equation}
S_Q= - \int d \mu \, Q(\delta \phi, \delta \theta) \log Q(\delta \phi, \delta \theta),
\end{equation} 
where $d\mu$ is the measure on the phase space of difference angles on the sphere.  To calculate this entropy, we coarse-grain the phase space so that the integral is transformed to a sum.  We expect the states delocalized in the chaotic sea to have large entropy by this measure, while those states well-localized around fixed points have low entropy.  This leaves some ambiguous situations, since highly excited states on regular tori also have high ``Husimi entropy''.  To improve the filter, we follow a procedure suggested by Korsch and coworkers \cite{Korsch}, which correlated the properties of the eigenstates to the classical phases space in order to distinguish the regular and irregular states for a nonlinear rotor.  In Fig. \ref{F4} we plot the values of $S_Q$ and $\langle J_z \rangle$.  The latter quantity correlates to the mean value of $\delta \theta$ in the semiclassical limit.  We see four distinct features in this plot.  Two lines of states with near constant $\langle J_z \rangle$ but increasing $S_Q$, boxed in Figs. (\ref{F4}a,b), correspond to the series of states localized around fixed points with increasing excitation (Figs \ref{F3}a,b).  The line of states with near constant $S_Q$ and increasing values of $\langle J_z \rangle$ , boxed in Fig. \ref{F4}c,  correspond to the series of states localized around the stable ``south pole" (Fig. \ref{F3}c).  Finally, the cluster of states with high values of both $S_Q$ and $\langle J_z \rangle$, boxed in Fig. \ref{F4}d,  correspond to the states delocalized in the chaotic sea that are concentrated near the original unstable fixed point at the ``north pole" of the regular dynamics.  There is no clean division between this cluster and states clearly localized on invariant tori.  A qualitative examination, denoted in Fig \ref{F4}, nonetheless gives us an indication of the chaotic subspace for these mixed dynamics.  Such an identification is useful for giving quantitative prediction of the dynamically generated entanglement, as we discuss in the next section.

\section{Entanglement}
\subsection{Calculating Entanglement}
We consider only pure states of the bipartite system.  Entanglement is then uniquely determined by the coefficients in the Schmidt decomposition of the joint state of the system,
\begin{equation}
\ket{\Psi}_{IJ}=\sum_{i} \sqrt{\lambda_i} \ket{u_i}_I \ket{v_i}_J ,
\end{equation}
where $\lambda_i$ are the eigenvalues of the reduced density matrix of either subsystem, and the Schmidt basis vectors $\{\ket{u_i}_I, \ket{v_i}_J\}$ are their respective eigenvectors. The entanglement $E$ is the Shannon entropy of the Schmidt coefficients,
\begin{equation}
E = -\sum_i \lambda_i  \log( \lambda_i).
\end{equation}
Determination of the Schmidt decomposition is typically a nontrivial task, requiring partial trace and diagonalization of the reduced density operator.  The Schmidt basis will generally depend on the state $\ket{\Psi}_{IJ}$.  For the system at hand, we have a unique situation -- within a subspace with a fixed value of $F_z$, the uncoupled basis of angular momentum is the Schmidt basis, {\em independent of the state}, as seen, e.g., Eq. (\ref{Eq:Eigen}).  Thus, for states within such subspaces, the entanglement is easily calculated as the Shannon entropy of the probability distribution of the state when expanded in the standard product basis. This not only simplifies calculations, but connects entanglement with the entropy of random states with respect to a fixed basis \cite{Wootters}.  

Throughout this section, we consider the $F_z=0$ subspace, and take $I=J=150$, corresponding to a Hilbert space of dimension $d=301$.  The maximum possible entanglement in this case is $E_{max} = \log d \approx 5.71$.

\subsection{Numerical Solutions}\label{S1}
\begin{figure}
\includegraphics[width=\linewidth]{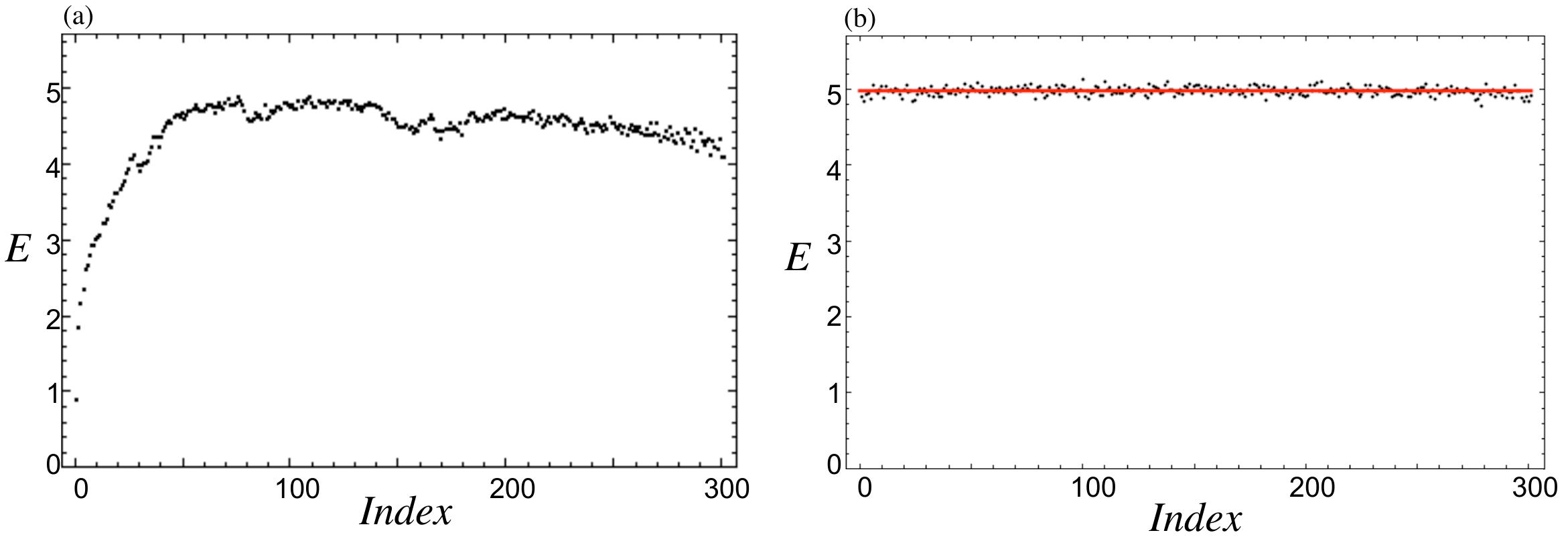}
\caption{Entanglement of the Floquet eigenstates. (a) Map corresponding to a mixed phase space: $\alpha=3/2,\beta=\pi/2$. (b) Map corresponding to global chaos: $\alpha=6,\beta=\pi/2$. The solid line gives the value expected from random matrix theory, Eq. (\ref{Eq:OER}).}
\label{F5}
\end{figure}

The entanglement of the Floquet eigenstates is easily calculated based on the discussion above.  Since the eigenstates reside in a subspace with fixed $F_z$, the uncoupled representation of angular momentum is the Schmidt basis, and the entanglement in a given eigenstate $\ket{k}$ is the Shannon entropy of the probability distribution of the expansion $\lambda^{(k)}_{m_J}= |c^{(k)}_{m_J}|^2$ from Eq. (\ref{Eq:Eigen}).  Figure \ref{F5} shows a list plot of this entanglement for a mixed phase space (as shown in Fig. (\ref{F2}b)) and  acompletely chaotic space (as shown in Fig. (\ref{F2}c)).  In the latter case, the entanglement values are clustered around the value expected from random matrix theory, discussed below.

Our main interest is to study the dynamically generated entanglement and its correlation with the classical phase space.  We wish to associate quantum states with our classical initial conditions. The ``most classical" state of a quantum system is a coherent state, so it would be natural to associate a point in our four-dimensional classical phase space with a product of spin coherent states.  These states, however, have support on several subspaces with different values of $F_z$, and thus correspond to a distribution of classical surfaces of sections.  To avoid this complication, we project our coherent states into the $M_F=0$ subspace, and then renormalize them, as described in Eq. (\ref{Eq:Project}).  This gives us a pure state, which though no longer separable, typically has a low entanglement and is localized around a point in the classical phase space in the relevant difference angles.

\begin{figure}
\includegraphics[width=\linewidth]{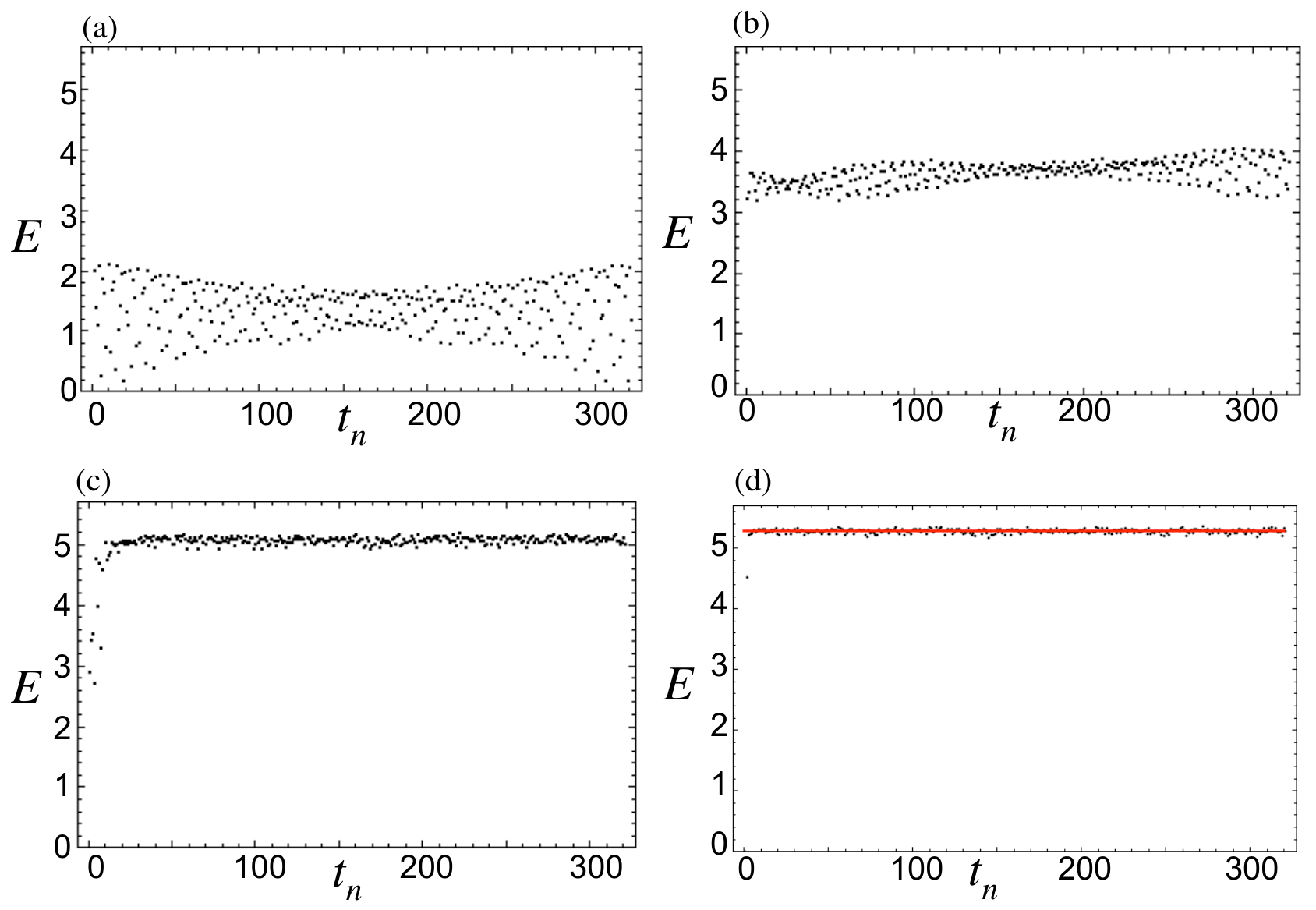}
\caption{ Dynamically generated entanglement as a function of the number of applications of the Floquet map. (a) Mixed phase space ($\alpha=3/2,\beta=\pi/2$), regular initial condition: $\ket{\psi_0}=\ket{I, I}\ket{J,-J}$. (b) Mixed phase space ($\alpha=3/2,\beta=\pi/2$), regular initial condition: $\ket{\psi_0}=\ket{\delta\theta=\pi/10,\delta\phi=53\pi/30}$. (c) Mixed phase space ($\alpha=3/2,\beta=\pi/2$), chaotic initial condition: $\ket{\psi_0}=\ket{I, -I}\ket{J,J}$ (d) Globally chaotic phase space ($\alpha=6,\beta=\pi/2$), chaotic initial condition:  $\ket{\psi_0}=\ket{\delta\theta=\pi/2,\delta\phi=\pi/3}$.  The solid line gives the value expected from random states in the Hilbert space, Eq. (\ref{Eq:UER}).}
\label{F6}
\end{figure}

The time-evolved state after $n$ applications of the Floquet operator to the projected coherent state is
\begin{equation}
\ket{\psi_n (\delta \theta, \delta \phi)}=U_{\tau}^n \ket{\delta \theta, \delta \phi} = \sum_k a_k e^{-i n\phi_k} \ket{k},
\end{equation}
expanded in the Floquet eigenstates, where $a_k = \braket{k}{\delta \theta, \delta \phi}$ is the initial spectral decomposition.  The Schmidt coefficients are the expansion of this state in the angular momentum product basis (the Schmidt basis) giving,
\begin{equation}
\label{Eq:lambda}
\lambda^{(n)}_{m_J}= \left|  \sum_k a_k   e^{-i n \phi_k} c^{(k)}_{m_J} \right|^2.
\end{equation}
according to Eqs. (\ref{Eq:Eigen}, \ref{Eq:lambda}). The Shannon entropy of these coefficients gives the dynamically evolved entanglement.  Figure \ref{F6} shows this quantum evolution for parameters such that the classical evolution is described by a mixed phase space.  For a coherent state initial condition chosen in the middle of a regular island ($\ket{\psi_0}=\ket{I, I}\ket{J,-J}=\ket{\delta\theta=-\pi,\delta\phi=0}$), the entanglement rises slowly and oscillates between high and low values.  For an initial condition in the chaotic sea ($\ket{\psi_0}=\ket{\delta\theta=\pi/2,\delta\phi=\pi/3}$), the entanglement rapidly rises and saturates to a near constant value, with small fluctuations about the steady state.

\begin{figure}
\includegraphics[width=\linewidth]{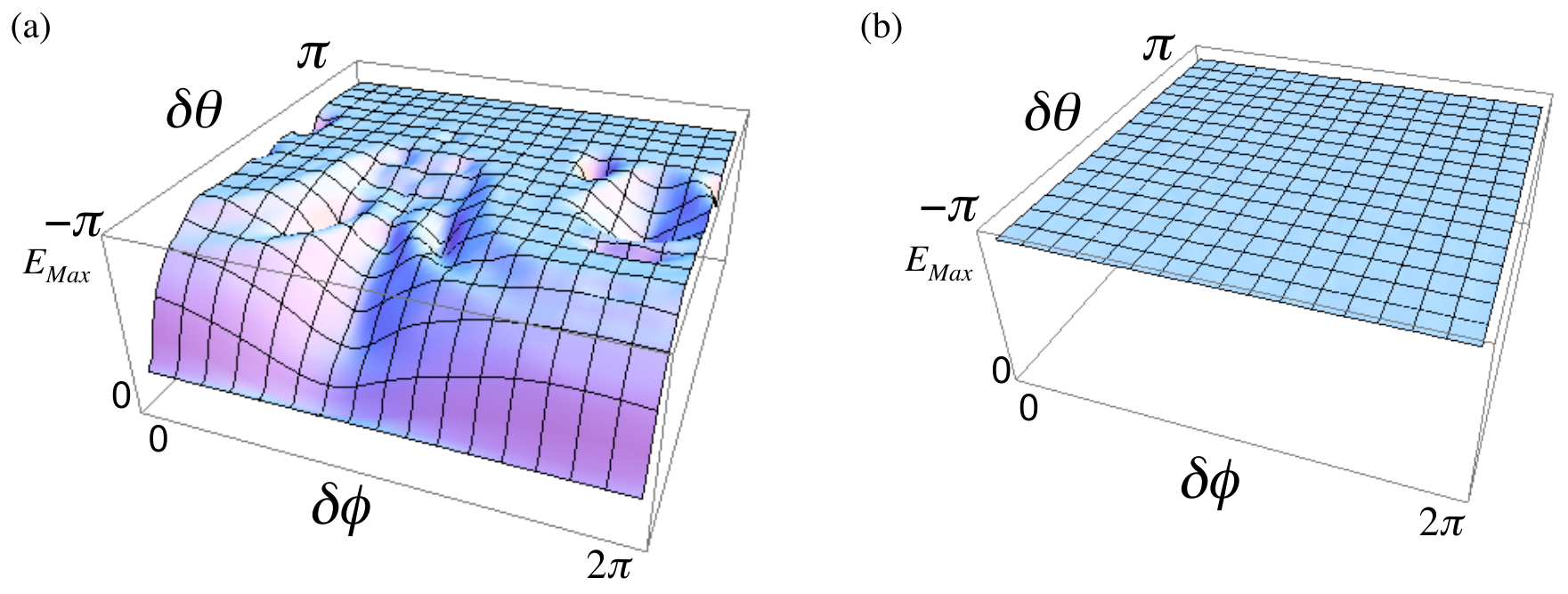}
\caption{Long-time average entanglement as a function of mean coordinate of the initial projected coherent state. (a) Mixed phase space: $\alpha=3/2,\beta=\pi/2$  (b) Globally chaotic phase space: $\alpha=6,\beta=\pi/2$)}
\label{F7}
\end{figure}

\begin{figure}
\includegraphics[width=\linewidth]{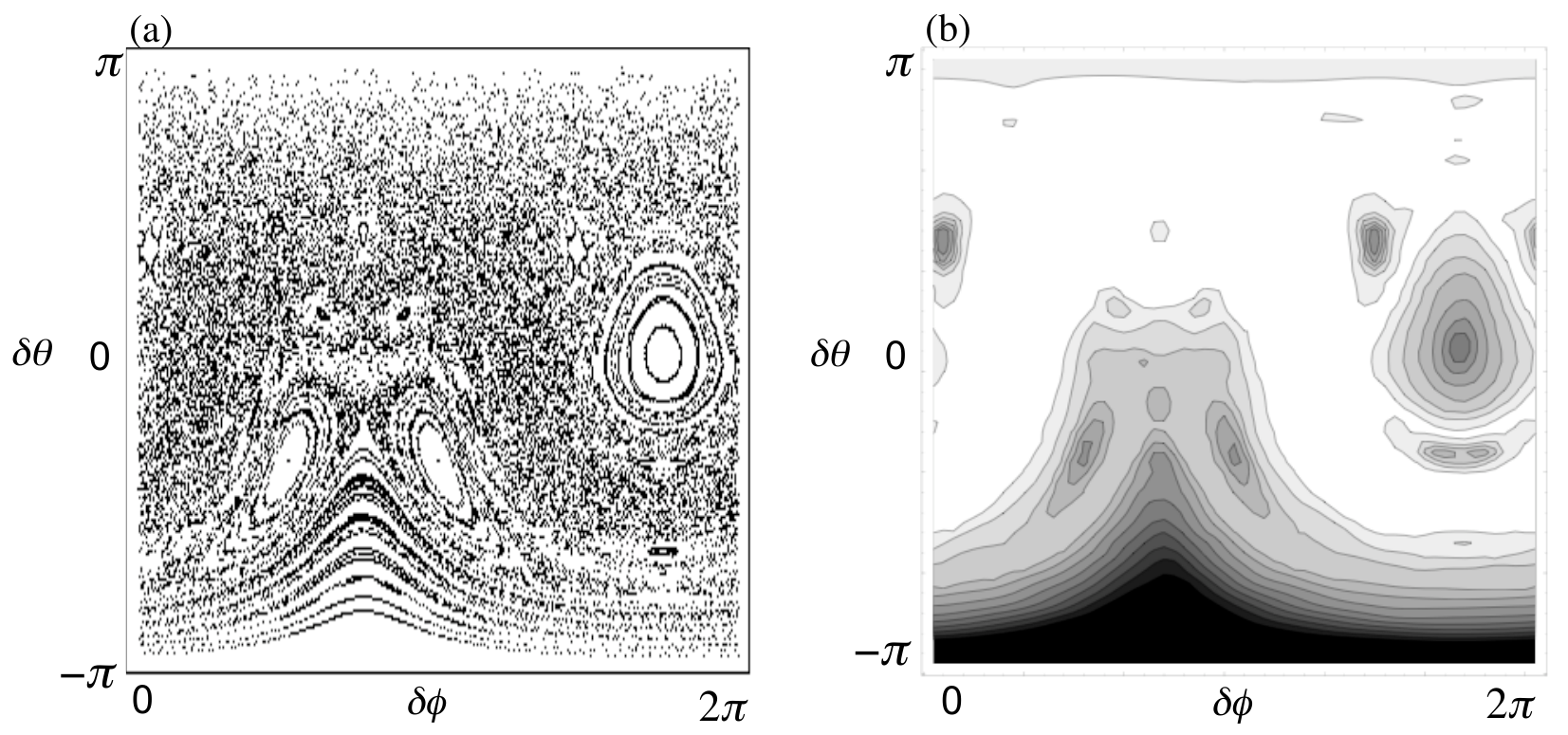}
\caption{Side-by-side comparison, showing dynamically generated entanglement as superb signature of classical chaos in a mixed phase space ($\alpha=3/2,\beta=\pi/2$). (a) Classical phase space, Poincar\'{e} section. (b) Long-time average entanglement as a function of mean coordinate of the initial projected coherent state}
\label{F8}
\end{figure}

In order to better explore how the entanglement evolution saturates to a particular value, we average over many time steps to find a long-time average of entanglement. We drop the first three hundred steps in order to remove transient effects and insure that the dynamics settle into a steady state, and then average over times steps 300-320.  By looking at a plot of this average, we can see how it correlates with initial conditions, a procedure initially carried out for the kicked top Hamiltonian by Wang {\em et al.} \cite{Wang}.  Figure \ref{F8} shows remarkably strong correlation between structures in the classical mixed phase space and the long-time entanglement average plot.  Chaotic initial conditions generally go to a higher average value than regular initial conditions, with the smallest values of entanglement generation near the classical fixed points.  Additionally, all initial conditions in the chaotic sea saturate to nearly the same average entanglement. 

For parameters corresponding to global chaos, we can see that the surface plot is very flat (see Fig. \ref{F8}a), with all initial conditions converging to nearly the same long-time entanglement average.  For the parameters at hand, averaging over all initial conditions, the dynamically generated entanglement is $\bar{E}_{dynam}=5.28$, as compared to the value $\bar{E}_{eigens}=4.97$ found for the average entanglement of the eigenstates of the Floquet map.   For the mixed phase space, the value of long-time entanglement is flat for initial conditions that correlate with the classical chaotic sea.  To find the entanglement characteristic of the chaotic initial conditions, we take a grid of coherent states across the phase space.  Each point in the grid is determined as ``regular" or ``chaotic" by the local Lyapunov exponent of the classical dynamics.  For those states with positive Lyapunov exponent we evolve according to the Floquet operator and calculate the long-time entanglement average, as described above.  Weighting these values according to the measure on phase space gives us an average entanglement of $\bar{E}_{dynam}=5.08$ in the chaotic sea, significantly lower than that for the globally chaotic phase space. Below, we interpret these results with statistics of random states in Hilbert space and their connection to quantum chaos.

\subsection{Entanglement and random states in Hilbert space}
The numerical studies in Sec. \ref{S1} reveal some empirical facts.  When the Floquet map corresponds to a fully chaotic phase space, the entanglement of the eigenstates are all nearly equal, with an average value independent of the coupling strength and below the maximum possible entanglement for the bipartite system.  Moreover, the dynamically generated entanglement when starting from a projected coherent state localized in a chaotic sea saturates to a nearly constant value after a few applications of the Floquet map.  In a mixed phase space, the amount of entanglement increases as the size of the chaotic sea increases.  For a completely chaotic space, the value no longer changes with coupling strength.  This saturation value is {\em different} from the entanglement seen in the eigenstates.  These facts leads us to conclude that the value of entanglement generation for chaotic maps is {\em statistical} in nature, as emphasized by Bandyopadhyay and Lakshminarayan \cite{Lakshminarayan/Bandyopadhyay2002,Lakshminarayan/Bandyopadhyay2004}, and Scott and Caves \cite{Scott/Caves}. The predicted values follow from the theory of random matrices and random states in Hilbert space, which we briefly review.

The random matrix conjecture of quantum chaos states that when the Hamiltonian (in an autonomous system) or Floquet map (in a periodically driven system) classically generates global chaos, the quantum operators have many of the statistical properties of a random matrix drawn from an appropriate ensemble depending on fundamental symmetries \cite{Haake}.  Systems with time-reversal symmetry (and no Kramer's degeneracy), invariant under orthogonal transformations, are described by random Hamiltonians picked from the Gaussian Orthogonal Ensemble of symmetric matrices (GOE) or Floquet maps from the Circular Orthogonal Ensemble of orthogonal matrices (COE).  Without time-reversal symmetry, the system is invariant only under general unitaries on the complex Hilbert space; the Hamiltonians and Floquet maps are random matrices chosen from the  Gaussian Unitary Ensemble (GUE) and Circular Unitary Ensemble (CUE), respectively.   This leads to the celebrated studies of level statistics for chaotic systems with different symmetries \cite{Haake}.  

Beyond the eigenvalues, the eigenvectors of these random operators have well defined statistical properties \cite{Kus88}.  In fact, both the Gaussian and Circular ensembles share the same eigenvector statistics.  To see this, note that since the random matrix ensembles are invariant under a group of transformations (orthogonal or complex unitary), the eigenvectors must {\em uniformly distributed} on a vector space according to the Haar measure that is invariant under that group.  The eigenvectors of random matrices are thus random states in a real or complex Hilbert space, picked according to the appropriate measure, as discussed by Wootters \cite{Wootters}. When discussing random state ensembles, we drop the ``G" and ``C" denotations.

To construct a Haar measure for sampling random states, we employ a parameterization equivalent to the Hurwitz parameterization of random unitaries \cite{Haar}. A measure can be constructed by connecting the space of unitaries with a manifold upon which there is a known geometric measure. A normalized state in an $d$-dimensional complex Hilbert space can be visualized as a point on the surface of a hypersphere in a $2d$ dimensional real space, where for each of the $d$ basis vectors in Hilbert space we assign a pair of orthogonal directions that project out the real and imaginary parts of the state's probability amplitude. The surface area of a differential patch on a hypersphere is then the probability measure for picking uniformly distributed random states.  The coordinates of a state, parameterized by angles on the hypersphere, and the corresponding measure over the space are
\begin{subequations}
\label{Eq:UE}
\begin{gather}
c_{1,r}=\cos\theta_1 ,\\
c_{1,i}=\sin\theta_1\cos\theta_2 ,\\
c_{n,r}=\sin\theta_1\dots\sin\theta_{2n-2}\cos\theta_{2n-1} ,\\ 
c_{n,i}=\sin\theta_1\dots\sin\theta_{2n-1}\cos\theta_{2n} ,\\ 
c_{d,r}=\sin\theta_1\dots\sin\theta_{2d-2}\cos\theta_{2d-1} ,\\
c_{d,i}=\sin\theta_1\dots\sin\theta_{2d-1} ,\\
d\lambda=N \sin^{2d-2}\theta_1\sin^{2d-3}\theta_2\dots \nonumber ,\\
\sin\theta_{2d-2} d\theta_1 d\theta_2\dots d\theta_{2d-1},
\end{gather}
\end{subequations} 
where $c_{n,r}$, $c_{n,i}$ are the real and imaginary expansion coefficients in the $n^{th}$ basis state, $d\lambda$ is the surface element, and $N$ is a normalization constant. The angles all range from $(0,\pi)$ except for the last angle which varies from $(0,2\pi)$. This defines the measure for random states in the unitary ensemble (UE).

For random states with probability that is invariant under an orthogonal transformation, the expansion coefficients can be chosen to be real. The probability measure is then the area element on $d$ dimensional hypersphere in a real space, with each direction corresponding to a basis vector of the Hilbert space . In this case the coordinates of the state and measure over the space are
\begin{subequations}
\label{Eq:OE}
\begin{gather}
c_{1,r}=\cos\theta_1 ,\\ 
c_{n,r}=\sin\theta_1\dots\sin(\theta_{n-1})\cos(\theta_{n}) ,\\ 
c_{d,r}=\sin\theta_1\dots\sin(\theta_{d-1}) ,\\
d\lambda=N \sin^{d-2}\theta_1\sin^{d-3}\theta_2\dots \nonumber ,\\
\sin\theta_{d-2}d \theta_1 d \theta_2\dots d \theta_{2d-1}.
\end{gather}
\end{subequations} 
This defines the measure for random states in the orthogonal ensemble (OE).

With these measures in hand, we can calculate expected values of entanglement of random states in an appropriate ensemble and compare them to the numerically predicted results.  For large $d$-dimensional spaces, the variance scales as $1/\sqrt{d}$  \cite{Wootters}, so when the states in question are well-described by the statistics above, we expect the expectation value to give good predictive power.  A well known example is the entanglement of a ``typical state" picked at random from a $d_1 \otimes d_2$ tensor product Hilbert space, with no other restrictions of symmetry.  The UE Haar measure average of the entanglement over the whole space gives \cite{Page,Scott/Caves, Hayden}
\begin{equation}
\bar{E}_{d_1 \otimes d_2}=\sum^{d_1d_2}_{k=d_1+1} \frac{1}{k}-\frac{d_1-1}{2d_2}, d_2\geq d_1.
\end{equation}
For large dimensions, $\bar{E}_{d_1 \otimes d_2} \approx \log d_1 - d_1/(2 d_2)$, which is close to the maximum possible value of entanglement, but saturates slightly below.  Typical pure states in an unconstrained bipartite Hilbert space are highly entangled \cite{Hayden}.  For the case at hand, symmetries constrain the accessible Hilbert space.  We thus turn to study the typical entanglement expected under these conditions.

\subsubsection{Typical entanglement in a subspace}\label{S2}
Our system has an additional symmetry, its rotational invariance around the $z$-axis.  This restricts our system so that eigenstates and dynamics take place in subspaces with fixed values of $F_z$.  Calculation of entanglement within a subspace is generally a nontrivial task as there is no simple expression for the entanglement in terms of variables that we can average over the Haar measure. In our case, there is a happy accident -- the uncoupled basis of angular momentum, $\vert J, m_J \rangle \otimes \vert I, M_F-m_J \rangle$, is also the Schmidt basis for {\em all} states in the subspace. This implies that we can take the {\em fixed} Schmidt vectors as the directions that define the space on a hypersphere, and thereby employ the same parameterization of the Haar measures as in Eqs. (\ref{Eq:UE},\ref{Eq:OE}), where now $d$ is the dimension of the subspace.  Note, this would not in general be possible for an arbitrary subspace because the entanglement is not a simple function of the expansion coefficients in a fixed basis. 

For a state in a fixed $F_z$ subspace, expanded in the uncoupled basis, $\ket{\Psi}=\sum c_{m_J} \ket{m_J} \ket{-m_J}$, the entanglement is
\begin{equation}
\label{Eq:Entropy}
E = - \sum_{m_J} \left| c_{m_J} \right|^2 \log \left(  \left| c_{m_J} \right|^2 \right ).
\end{equation}
For random states in the subspace picked according to the orthogonally invariant Haar measure, the coefficients are taken to be real and distributed on the hypersphere according to Eq. (\ref{Eq:OE}). The contribution of each term in the expression for the entanglement given above should be equal, so we can shortcut by integrating only the first term, and multiplying by the number of terms, $d$. We normalize by an integral over the measure for that variable.  The result for the orthogonal ensemble is
\begin{eqnarray}
\label{Eq:OER}
\bar{E}_{OE} &=&d\frac{ - \int \left| \cos\theta_1 \right|^2 \log \left(  \left| \cos\theta_1 \right|^2 \right ) \sin^{d-2} \theta_1 d \theta_1}{ \int \sin^{d-2}\theta_1 d \theta_1} \nonumber \\
&=&\mathcal{H}_{d/2}+\log 4-2,
\end{eqnarray}
where
\begin{equation}
\mathcal{H}_D = 1 + 1/2 + 1/3 + \dots + 1/D
\end{equation}
is the harmonic series. 

For the states picked according to the unitarily invariant Haar measure within this subspace, it is useful to first simplify our parameterization by specifying the magnitudes of the expansion coefficients in terms of the angles on the hypersphere, rather than the real and imaginary parts of the expansion coefficients. Our new parameterization and the associated surface element are as follows:
\begin{subequations}
\begin{gather}
\left|c_{1}\right|=\cos\theta_1 \\
\left|c_{m}\right|=\sin\theta_1\dots\sin\theta_{m-1}\cos\theta_m \\
\left|c_{d}\right|=\sin\theta_1\dots\sin\theta_{d-1} \\
d\lambda=N \sin^{2d-3}\theta_1\sin^{2d-5}\theta_2\dots\nonumber \\
\sin\theta_{d-1}\cos\theta_1\dots\cos\theta_{d-1}d\theta_1 d\theta_2...d\theta_{2d-1},
\end{gather}
\end{subequations} 
where $\theta_m$ now ranges from $(0,\pi/2)$. Since the entanglement for a state in the subspace depends only on the the magnitudes $\{|c_m|\}$, Eq. (\ref{Eq:Entropy}) can be expressed in terms of this parameterization of the manifold.  Performing the average, the typical entanglement for a state picked for the UE, restricted to a $F_z$ subspace, is
\begin{eqnarray}
\label{Eq:UER}
\bar{E}_{UE} &=&d\frac{ -  \int \left| \cos\theta_1 \right|^2 \log \left(  \left| \cos\theta_1 \right|^2 \right ) \sin^{2d-3} \theta_1 \cos\theta_1 d \theta_1}{ \int \sin^{2d-3}\theta_1 \cos\theta_1 d \theta_1} \nonumber \\
&=& \mathcal{H}_d-1.
\end{eqnarray}
These averages hold regardless of dimension of the space, though the variance of the distribution rapidly narrows as $d$ increases.

In the limit of large dimensional spaces, 
\begin{subequations}
\label{Eq:Log}
\begin{gather}
\bar{E}_{\text{OE}} \rightarrow \log d - 2 + \log 2 + \gamma ,\\
\bar{E}_{\text{UE}} \rightarrow \log d - 1 + \gamma,
\end{gather}
\end{subequations}
where $\gamma \approx 0.577 $ is Euler's constant.  These entanglement values are equal to the entropy of a random state in a real or complex Hilbert space with respect to a fixed basis, as discussed by Wootters \cite{Wootters} and Zyczkowski \cite{Zyczkowski}.  This is not surprising since the Schmidt basis for our system is fixed when the state is confined to a subspace with fixed $F_z$. Generally, this is not true, and the entropy of the squared expansion coefficients with respect to a basis is not equal to the typical entanglement.  For example, for the full tensor product space, for large dimensional Hilbert spaces with $d_1=d_2$, $\bar{E}_{d_1 \otimes d_2} \rightarrow \log d_1 -1/2$, which differs from the Wooters/Zyczkowski entropy, Eq. (\ref{Eq:Log}), taking $d=d_1^2$.

As an aside, we can repeat our calculations for the linear entropy, an entanglement monotone. The linear entropy is determined by the purity of the reduced density operator of one subsystem,
\begin{equation}
S_{\text{L}}(\rho)=1-Tr(\rho_{\text{red}}^2) = 1- \sum_m \lambda_m^2  =1-\sum_m \left| c_m \right|^4
\end{equation}
where $\lambda_m = \left| c_m \right|^2$ are the Schmidt coefficients for a state in the subspace.  We repeat our integrals over the appropriate manifolds and find 
\begin{equation}
\bar{S}_{\text{L,OE}}=1-\frac{3}{d+2}, \hspace{1 pc} \bar{S}_{\text{L,UE}}=1-\frac{2}{d+1},
\end{equation}
the same results found by Brown and Viola by different methods \cite{Brown/Viola}.

\subsubsection{Typical entanglement prediction for the kicked coupled-tops}
With the results of Sec. \ref{S2} in hand, we can compare the predictions of the typical entanglement of random states to the entanglement found numerically in Sec. \ref{S1}.  Since the system is time reversal invariant without Kramer's degeneracy as shown in Sec. \ref{S3}, under the random matrix conjecture of quantum chaos, we expect the eigenstates of the Floquet operator for globally chaotic classical dynamics to be random states chosen from the OE.  The eigenstates are restricted to a subspace with fixed value of $F_z$, so Eq. (\ref{Eq:OER}) applies.  We consider the $F_z=0$ subspace with dimension $d=2J+1$.  For spin $J=150$, one finds $\bar{E}_{\text{OE}} = 4.98$, in excellent agreement with the mean entanglement of the eigenstates for the globally chaotic case, $\bar{E}_{\text{eigens}}=4.97$.  

Next we consider the dynamically generated entanglement, starting from a spin coherent product state projected into the $F_z=0$ subpace.  The key conjecture, seen numerically in prior studies, is that chaotic maps acting on a fiducial state generate states with the statistics of random states in Hilbert space, chosen according to the appropriate ensemble.  However, contrary to prior claims \cite{Lakshminarayan/Bandyopadhyay2002}, though the Floquet operator is a member of the COE, the dynamically generated state is {\em not} random according to the OE.  To see this, first note that since the Floquet operator is a member of the COE, we know the eigenstates are time-reversal invariant, $T\ket{k}=\ket{k}$.  However, according to Eq. (\ref{Eq:Reversal}), time reversal acting on the dynamically evolved state gives
\begin{equation}
T\ket{\psi_n (\delta \theta, \delta \phi)}=\sum_k a_k^* e^{+i n\phi_k} \ket{k} \neq \ket{\psi_{n} (\delta \theta, \delta \phi)}.
\end{equation}
Thus, the dynamically evolved state is {\em not} an eigenstate of the time reversal operator.  This is true even when the initial state itself is a time-reversal eigenstate (e.g., the coherent state at the pole, $\ket{\psi(0)} = \ket{I,m_I=-m_J}\ket{J,-m_J}$), in which case $T\ket{\psi_n}=\ket{\psi_{-n}}$.  Thus the appropriate ensemble here is the UE, and we expect the dynamically generated entanglement to be predicted by random states with these statistics.  This is indeed born out in the numerics.  For the globally chaotic map, we evolve and average to find the steady state value, as discussed in Sec. \ref{S1}. The long-time entanglement average is almost independent of the initial coherent state, projected in the $F_z=0$ subspace. For these initial condition Eq. (\ref{Eq:UER}) predicts $\bar{E}_{UE}=5.28$ in good agreement with the long-time average value of 5.28. 

In the case of a mixed phase space, we saw that the long-time entanglement average was almost constant for initial states localized in the chaotic sea.  Clearly, this value of entanglement is a statistical property of Hilbert space.  Just as the quantum dynamics lead to a random state in the entire $F_z=0$ subspace when the classical dynamics are globally chaotic, for a classically mixed phase space, based on Percival's conjecture, the quantum dynamics generate a random state in the {\em chaotic subspace}. The structure of the chaotic sea cannot be described by a simple symmetry, so we cannot determine the entanglement of a typical state analytically. However, we can filter the eigenstates to determine which are in the chaotic subspace, as discussed in Sec. \ref{S3}, and sample randomly from a unitarily invariant measure over this subspace in order to find the typical entanglement value.  In this case there is no simple expression for the entanglement as a function of the states, so we cannot analytically take the average over the appropriate measure as before. Instead, we generate a large number of random states in the chaotic subspace, and find their entanglements. We do this by picking the real and imaginary parts of the expansion coefficients with respect to the chaotic eigenstates according to a Gaussian distribution. After normalizing, the entanglement is calculated for this state, and the process is repeated 100 times. The results are averaged to find an estimate of the average entanglement of a random state in the chaotic sea. We find that the average entanglement of a random state in the chaotic subspace picked according to the UE distribution is 5.13, in good agreement with the numerically determined value of $\bar{E}_{dynam}=5.08$ found in Sec. \ref{S1}.   Part of this discrepancy is likely due to the greater degree of variation of entanglement across the chaotic sea in the mixed phase space compared to the relatively flat completely chaotic phase space.   In addition our filter for determining the members of the chaotic subspace was somewhat crude with an ambiguous ``grey zone".  We would expect this to improve deeper in the semiclassical regime, where Percival's conjecture applies better.  

\section{Discussion and Summary}
Classical chaotic dynamics lead to ergodic mixing in phase space. Quantum analogs of ergodicity have long been considered, including ``spectral chaos" \cite{Heller84} and increase in entropy associated with the wave function when expanded in a fixed (non-stationary-state) basis \cite{Peres}.  Recent numerical studies indicate that quantum dynamics generated by nonintegrable Hamiltonians are ergodic in the sense that they generate pseudo-random states in a Hilbert space, chosen according to a Haar measure that is dictated by the symmetries of the system \cite{Scott/Caves, Emerson}.  Such a result is not new, having its roots in the random matrix theory conjecture of quantum chaos \cite{Bohigas} -- the typical Hamiltonian of a nonintegrable system has the statistical properties of random matrices of an ensemble picked according to the symmetries of the system under time reversal.  The classic works on the subject, however, focus on the properties of the stationary states and spectra -- Berry's ``quantum chaology" \cite{Berry87}.  Dynamical ergodicity in quantum mechanics has been harder to pin down and is still a subject of some controversy.

The existence of quantum dynamical ergodicity has implications for the dynamical generation of entanglement.  It is well known that for large dimensional bipartite Hilbert spaces, a random state is highly entangled with almost the maximum entanglement allowed by the dimension \cite{Hayden}.  As the large dimensional limit is equivalent to the $\hbar \rightarrow 0$ semiclassical limit, and to the degree that the quantum analogs of chaotic Hamiltonians generate random states, one expects near maximal dynamical generation of entanglement in quantum chaos, to a value that is predicted by the statistics at hand.  This is not to say that regular dynamics (quantum analogs of integrable motion) cannot lead to highly entangled states.  Indeed, such behavior is seen, and has been previously noted in \cite{Kus04}.  Regular dynamics, however, show oscillatory behavior, including in the generation of entanglement.  Chaotic dynamics, by contrast, lead to quasi-steady state behavior, and typically lead to higher values of time-averaged entanglement than regular motion.  Taken together, these facts imply that the long-time average entanglement in a bipartite system should be a strong signature of classical chaos, closely associated with ergodicity in the two dynamical descriptions.

We have studied the relationship between entanglement and chaos for a system of isotropically coupled tops, in which one of the tops receives a periodic kick around a fixed axis.  In contrast to most previous studies, here the chaos and entanglement arise from the same coupling mechanism.  This allows us to remove ambiguities that have been discussed regarding the effect of coupling between the subsystems vs. the chaoticity in the individual subsystems.  Moreover, the rotational symmetry of the system allows us to easily calculate and interpret the entanglement because the Schmidt basis, within a subspace defined by a fixed value of the conserved angular momentum, is independent of the joint state of the system. This unique property connects entanglement in our system to the entropy of the probability distribution of a random state with respect to a fixed basis.  

The results reported here give further confirmation to the fact that chaotic systems take quantum initial conditions to pseudo-random quantum states, and that the high long-time entanglement average of states undergoing quantum chaotic dynamics is just that of a typical state in the Hilbert space. We see the confirmation of this picture in the excellent agreement between the properties of ensembles of quantum states and the numerical results for the eigenvector statistics and long-time entanglement average for the completely chaotic system. This approach was also found to be highly flexible, applying to subspaces and mixed phase spaces.

For parameters with global chaos, independent of the initial condition, the dynamically generated entanglement is well-predicted by a random state of the appropriate Haar-averaged unitary ensemble. This value is distinct from the expected value of entanglement in the orthogonal ensemble that describes the Floquet eigenstates for this time-reversal invariant system.  Dynamical evolution changes the measure over which we must average, even though the Floquet operator itself is a member of the COE, and even when the initial state itself is an eigenstate of time reversal.  We are able to derive exact analytic results for these expected values when the state is restricted to a subspace dictated by the symmetries. 

For a mixed phase space, numerical plots of the long-time entanglement average entanglement exhibit beautiful correlation with the regular islands and the chaotic sea.  Within the chaotic sea, the entanglement is fairly uniform, with a value characteristic of the ergodic mixing.  We show this by appealing to Percival's conjecture.  In the semiclassical limit, the Hilbert space approximately decomposes into subspaces spanned by the eigenstates that condense on the regular islands and those that are spread through the chaotic sea.  We determine the chaotic subspace by filtering the Floquet eigenstates according to their entropy in phase space and by the expected values of observables associated with the regular islands.  Given an identification of the chaotic subspace, we performed a Haar measure average of the expected entanglement and found reasonable agreement with the dynamically generated value for states initially localized in the chaotic sea.  There is still some deviation between our predictions and the numerical results, but we believe that the basic reasoning behind this method is sound. For systems in which symmetry more cleanly separates the chaotic and regular subspaces \cite{Tomsovic}, we expect that the Haar measure average over the unitary group will give a very accurate prediction of the entanglement generation in the chaotic sea for mixed phase spaces.

\begin{acknowledgements}
This project has greatly benefited from long-time collaborations with Prof. Shohini Ghose and Prof. Arjendu Pattanayak, and with and students Parin Sripakdeevong and Leigh Norris, whose work on both classical and quantum dynamics of the system helped guide us. We also thank Andrew Scott, Steve Flammia, Seth Merkel and Andrew Silberfarb for numerous enlightening discussions.  This work was supported by a grant from the National Science Foundation, Award Number 0555573.
\end{acknowledgements}

\end{document}